\documentstyle[twoside,epsf]{article}

\catcode`\@=11
\long\def\@makefntext#1{ 
\protect\noindent \hbox to 3.2pt {\hskip-.9pt
$^{{\eightrm\@thefnmark}}$\hfil}#1\hfill} 

\def\thefootnote{\fnsymbol{footnote}}
 \def\@makefnmark{\hbox to 0pt{$^{\@thefnmark}$\hss}}  

\def\ps@myheadings{\let\@mkboth\@gobbletwo
\def\@oddhead{\hbox{} 
\rightmark\hfil\eightrm\thepage}
\def\@oddfoot{}\def\@evenhead{\eightrm\thepage\hfil 
\leftmark\hbox{}}\def\@evenfoot{}
\def\sectionmark##1{}\def\subsectionmark##1{}}

\oddsidemargin=\evensidemargin
\renewcommand{\thefootnote}{\fnsymbol{footnote}}

\newcounter{sectionc}\newcounter{subsectionc}\newcounter{subsubsectionc}
\renewcommand{\section}[1] {\vspace{12pt}\addtocounter{sectionc}{1}
\setcounter{subsectionc}{0}\setcounter{subsubsectionc}{0}\noindent
	{\tenbf\thesectionc. #1}\par\vspace{5pt}}
\renewcommand{\subsection}[1] {\vspace{12pt}\addtocounter{subsectionc}{1}
	\setcounter{subsubsectionc}{0}\noindent
	{\bf\thesectionc.\thesubsectionc. {\kern1pt \bfit #1}}\par\vspace{5pt}}
\renewcommand{\subsubsection}[1] {\vspace{12pt}\addtocounter{subsubsectionc}{1}
	\noindent{\tenrm\thesectionc.\thesubsectionc.\thesubsubsectionc.
	{\kern1pt \tenit #1}}\par\vspace{5pt}}
\newcommand{\nonumsection}[1] {\vspace{12pt}\noindent{\tenbf #1}
	\par\vspace{5pt}}

\newcounter{appendixc}
\newcounter{subappendixc}[appendixc]
\newcounter{subsubappendixc}[subappendixc]
\renewcommand{\thesubappendixc}{\Alph{appendixc}.\arabic{subappendixc}}
\renewcommand{\thesubsubappendixc}
	{\Alph{appendixc}.\arabic{subappendixc}.\arabic{subsubappendixc}}

\renewcommand{\appendix}[1] {\vspace{12pt}
        \refstepcounter{appendixc}
        \setcounter{figure}{0}
        \setcounter{table}{0}
        \setcounter{lemma}{0}
        \setcounter{theorem}{0}
        \setcounter{corollary}{0}
        \setcounter{definition}{0}
        \setcounter{equation}{0}
        \renewcommand{\thefigure}{\Alph{appendixc}.\arabic{figure}}
        \renewcommand{\thetable}{\Alph{appendixc}.\arabic{table}}
        \renewcommand{\theappendixc}{\Alph{appendixc}}
        \renewcommand{\thelemma}{\Alph{appendixc}.\arabic{lemma}}
        \renewcommand{\thetheorem}{\Alph{appendixc}.\arabic{theorem}}
        \renewcommand{\thedefinition}{\Alph{appendixc}.\arabic{definition}}
        \renewcommand{\thecorollary}{\Alph{appendixc}.\arabic{corollary}}
        \renewcommand{\theequation}{\Alph{appendixc}.\arabic{equation}}
        \noindent{\tenbf Appendix \theappendixc #1}\par\vspace{5pt}}
\newcommand{\subappendix}[1] {\vspace{12pt}
        \refstepcounter{subappendixc}
        \noindent{\bf Appendix \thesubappendixc. {\kern1pt \bfit #1}}
	\par\vspace{5pt}}
\newcommand{\subsubappendix}[1] {\vspace{12pt}
        \refstepcounter{subsubappendixc}
        \noindent{\rm Appendix \thesubsubappendixc. {\kern1pt \tenit #1}}
	\par\vspace{5pt}}

\newcommand{\textlineskip}{\baselineskip=13pt}
\newcommand{\smalllineskip}{\baselineskip=10pt}

\def\eightcirc{
\begin{picture}(0,0)
\put(4.4,1.8){\circle{6.5}}
\end{picture}}
\def\eightcopyright{\eightcirc\kern2.7pt\hbox{\eightrm c}}

\newcommand{\copyrightheading}[1]
	{\vspace*{-2.5cm}\smalllineskip{\flushleft
	{\eightrm Modern Physics Letters B, #1}\\
	{\eightrm $\eightcopyright$\, World Scientific Publishing
	 Company}\\
	 }}

\newcommand{\pub}[1]{{\begin{center}\eightrm\smalllineskip
	Received #1\\
	\end{center}
	}}

\def\abstracts#1#2#3{{
	\centering{\begin{minipage}{4.5in}\baselineskip=10pt\eightrm
	\centerline{ABSTRACT}
	\parindent=0pt #1\par
	\parindent=15pt #2\par
	\parindent=15pt #3
	\end{minipage} }\par}}

\newcommand{\bibit}{\nineit}
\newcommand{\bibbf}{\ninebf}
\renewenvironment{thebibliography}[1]			
	{\ninerm
	 \baselineskip=11pt				
	 \begin{list}{\arabic{enumi}.}
	{\usecounter{enumi}\setlength{\parsep}{0pt}
	 \setlength{\leftmargin 17pt}{\rightmargin 0pt}	
	 \setlength{\itemsep}{0pt} \settowidth		
	{\labelwidth}{#1.}\sloppy}}{\end{list}}

\newcounter{itemlistc}
\newcounter{romanlistc}
\newcounter{alphlistc}
\newcounter{arabiclistc}

\newcommand{\fcaption}[1]{
        \refstepcounter{figure}
        \setbox\@tempboxa = \hbox{\eightrm Fig.~\thefigure. #1}
        \ifdim \wd\@tempboxa > 5in
           {\begin{center}
        \parbox{5in}{\eightrm \smalllineskip Fig.~\thefigure. #1 }
            \end{center}}
        \else
             {\begin{center}
             {\eightrm Fig.~\thefigure. #1}
              \end{center}}
        \fi}

\newcommand{\tcaption}[1]{
        \refstepcounter{table}
        \setbox\@tempboxa = \hbox{\eightrm Table~\thetable. #1}
        \ifdim \wd\@tempboxa > 5in
           {\begin{center}
        \parbox{5in}{\eightrm\smalllineskip Table~\thetable. #1 }
            \end{center}}
        \else
             {\begin{center}
             {\eightrm Table~\thetable. #1}
              \end{center}}
        \fi}

\def\@citex[#1]#2{\if@filesw\immediate\write\@auxout	
	{\string\citation{#2}}\fi			
\def\@citea{}\@cite{\@for\@citeb:=#2\do			
	{\@citea\def\@citea{,}\@ifundefined		
	{b@\@citeb}{{\bf ?}\@warning
	{Citation `\@citeb' on page \thepage \space undefined}}
	{\csname b@\@citeb\endcsname}}}{#1}}

\newif\if@cghi
\def\cite{\@cghitrue\@ifnextchar [{\@tempswatrue
	\@citex}{\@tempswafalse\@citex[]}}
\def\citelow{\@cghifalse\@ifnextchar [{\@tempswatrue
	\@citex}{\@tempswafalse\@citex[]}}
\def\@cite#1#2{{$\null^{#1}$\if@tempswa\typeout
	{IJCGA warning: optional citation argument
	ignored: `#2'} \fi}}

\def\pmb#1{\setbox0=\hbox{#1}
	\kern-.025em\copy0\kern-\wd0
	\kern.05em\copy0\kern-\wd0
	\kern-.025em\raise.0433em\box0}

\def\fnt#1#2{\footnotetext{\kern-.3em
	{$^{\mbox{\scriptsize #1}}$}{#2}}}

\def\fpage#1{\begingroup
\voffset=.3in
\thispagestyle{empty}\begin{table}[b]\centerline{\footnotesize #1}
	\end{table}\endgroup}

\def\runninghead#1#2{\pagestyle{myheadings}
\markboth{{\eightit{\quad #1}}\hfill}{\hfill{\eightit{#2\quad}}}}

\font\tenbf=cmbx10
\font\tenit=cmti10
\font\tenit=cmti10
\font\bfit=cmbxti10 at 10pt
 1
 1
 1
\font\ninebf=cmbx9
\font\ninerm=cmr9
\font\nineit=cmti9

\font\eightrm=cmr8
\font\eightit=cmti8

\def\qed{\hbox{${\vcenter{\vbox{                          
   \hrule height 0.4pt\hbox{\vrule width 0.4pt height 6pt
   \kern5pt\vrule width 0.4pt}\hrule height 0.4pt}}}$}}

\runninghead{C. W. J. Beenakker and M. Patra} {Photon Shot Noise}

\begin{document}
\newcommand{\case}[2]{\mbox{${\textstyle\frac{#1}{#2}}$}}
\newcommand{\openone}{\leavevmode\hbox{\small1\kern-3.3pt\normalsize1}}
\newcommand{\gtequiv}{\lower2pt\hbox{$\:\stackrel{>}{
\scriptstyle\sim}\:$}}
\newcommand{\ltequiv}{\lower2pt\hbox{$\:\stackrel{<}{
\scriptstyle\sim}\:$}}
\normalsize\textlineskip
{\thispagestyle{empty}
\setcounter{page}{1}
\renewcommand{\thefootnote}{\fnsymbol{footnote}} 
\copyrightheading{to be published as ``Brief Review''}
\vspace*{0.88truein}
\fpage{1}

\centerline{\bf PHOTON SHOT NOISE}
\vspace{0.37truein}
\centerline{\footnotesize C. W. J. Beenakker and M. Patra}
\vspace*{0.015truein}
\centerline{\footnotesize\it Instituut-Lorentz, Universiteit Leiden}
\baselineskip=10pt
\centerline{\footnotesize\it P.O. Box 9506, 2300 RA Leiden, The Netherlands}
\vspace{0.225truein}
\pub{May 1999}
\vspace*{0.21truein}

\abstracts{\noindent
A recent theory is reviewed for the shot noise of coherent radiation
propagating through a random medium. The Fano factor $P/\bar{I}$ (the ratio of
the noise power and the mean transmitted current) is related to the scattering
matrix of the medium. This is the optical analogue of B\"{u}ttiker's formula
for electronic shot noise. Scattering by itself has no effect on the Fano
factor, which remains equal to $1$ (as for a Poisson process). Absorption and
amplification both increase the Fano factor above the Poisson value. For strong
absorption $P/\bar{I}$ has the universal limit $1+\frac{3}{2}f$, with $f$ the
Bose-Einstein function at the frequency of the incident radiation. This is the
optical analogue of the one-third reduction factor of electronic shot noise in
diffusive conductors. In the amplifying case the Fano factor diverges at the
laser threshold, while the signal-to-noise ratio $\bar{I}^{2}/P$ reaches a
finite, universal limit.
}{}{}

\vspace*{-3pt}\textlineskip
\section{Introduction}
\noindent
Analogies in the behavior of photons and electrons provide a continuing source
of inspiration in mesoscopic physics.\cite{But99} Two familiar examples are the
analogies between weak localization of electrons and enhanced backscattering of
light and between conductance fluctuations and optical speckle.\cite{Alt91} The
basis for these analogies is the similarity between the single-electron
Schr\"{o}dinger equation and the Helmholtz equation. The Helmholtz equation is
a classical wave equation, and indeed the study of mesoscopic phenomena for
light has been limited mostly to {\em classical\/} optics. A common theme in
these studies is the interplay of interference and multiple scattering by
disorder. The extension to {\em quantum\/} optics adds the interplay with
vacuum fluctuations as a new ingredient.

Recently a theoretical approach to the quantum optics of disordered media was
proposed,\cite{Bee98} that utilizes the methods of the random-matrix theory of
quantum transport.\cite{Bee97,Guh98} The random matrix under consideration is
the scattering matrix. The basic result of Ref.\ 3 is a relationship
between the scattering matrix and the photocount distribution. It was applied
there to the statistics of blackbody radiation and amplified spontaneous
emission. This work was reviewed in Ref.\ 6. Here we review a later
development,\cite{Pat99} the optical analogue of electronic shot noise.

\eject}
\textheight=7.8truein
\setcounter{footnote}{0}
\renewcommand{\thefootnote}{\alph{footnote}}

Shot noise is the time-dependent fluctuation of the current
$I(t)=\bar{I}+\delta I(t)$ (measured in units of particles/s) resulting from
the discreteness of the particles. The noise power
\begin{equation}
P=\int_{-\infty}^{\infty}dt\,\overline{\delta I(0)\delta I(t)} \label{Pdef}
\end{equation}
quantifies the size of the fluctuations. (The bar $\overline{\cdots}$ indicates
an average over many measurements on the same system.) For independent
particles the current fluctuations form a Poisson process, with power $P_{\rm
Poisson}=\bar{I}$ equal to the mean current. The ratio $P/P_{\rm Poisson}$
(called the Fano factor\cite{Fan46}) is a measure of the correlations between
the particles.

For electrons, correlations resulting from the Pauli exclusion principle reduce
$P$ below $P_{\rm Poisson}$. (See Ref.\ 9 for a review.) The ratio
$P/P_{\rm Poisson}$ is expressed in terms of traces of the transmission matrix
$t$ at the Fermi energy by\cite{But90}
\begin{equation}
P/P_{\rm Poisson}=1-\frac{{\rm Tr}\,(tt^{\dagger})^{2}}{{\rm
Tr}\,tt^{\dagger}}. \label{Pttdagger}
\end{equation}
This formula holds at zero temperature (no thermal noise). In the absence of
scattering all eigenvalues of the transmission-matrix product $tt^{\dagger}$
are equal to unity, hence $P=0$. This absence of shot noise is realized in a
ballistic point contact.\cite{Khl87,Les89} At the other extreme, in a tunnel
junction all transmission eigenvalues are $\ll 1$, hence $P=P_{\rm
Poisson}$.\cite{Sch18} A disordered metallic conductor is intermediate between
these two extremes, having $P=\frac{1}{3}P_{\rm Poisson}$.\cite{Bee92,Nag92}

For the optical analogue we consider a monochromatic laser beam (frequency
$\omega_{0}$) incident in a single mode (labelled $m_{0}$) on a waveguide
containing a disordered medium (at temperature $T$). The radiation from a laser
is in a coherent state. The photostatistics of coherent radiation is that of a
Poisson process,\cite{Man95} hence $P=P_{\rm Poisson}$ for the incident beam.
The question addressed in this work is: How does the ratio $P/P_{\rm Poisson}$
change as the radiation propagates through the random medium? We saw that, for
electrons, scattering increases this ratio. In contrast, in the optical
analogue scattering by itself has no effect: $P$ remains equal to $P_{\rm
Poisson}$ if the incident beam is only partially transmitted --- provided the
scattering matrix remains unitary. A non-unitary scattering matrix, resulting
from absorption or amplification of radiation by the medium, increases the
ratio $P/P_{\rm Poisson}$. This excess noise can be understood as the beating
of coherent radiation with vacuum fluctuations of the electromagnetic
field.\cite{Hen96}

Photon shot noise has been studied extensively in systems where the scattering
is one-dimensional (for example, randomly layered media).\cite{Jef93,Mat97} No
formula of the generality of Eq.\ (\ref{Pttdagger}) was needed for those
investigations. In order to go beyond the one-dimensional case, we have derived
the optical analogue of Eq.\ (\ref{Pttdagger}). The result is\cite{Pat99}
\begin{equation}
P/P_{\rm
Poisson}=1+2f(\omega_{0},T)\frac{[t^{\dagger}
(\openone-rr^{\dagger}-tt^{\dagger})t]_{m_{0}m_{0}}} {[t^{\dagger}t]_{m_{0}m_{0}}}, \label{Pexcess}
\end{equation}
where $f(\omega,T)=[\exp(\hbar\omega/kT)-1]^{-1}$ is the Bose-Einstein
function. Eq.\ (\ref{Pexcess}) contains both the transmission matrix $t$ and
the reflection matrix $r$ (evaluated at frequency $\omega_{0}$). For a unitary
scattering matrix, $rr^{\dagger}+tt^{\dagger}$ equals the unit matrix
$\openone$, hence the term proportional to $f$ in Eq.\ (\ref{Pexcess}) vanishes
and $P=P_{\rm Poisson}$. Absorption and amplification both lead to an
enhancement of $P$ above $P_{\rm Poisson}$. For an absorbing system the matrix
$\openone-rr^{\dagger}-tt^{\dagger}$ is positive definite and $f>0$, so
$P/P_{\rm Poisson}>1$. In an amplifying system
$\openone-rr^{\dagger}-tt^{\dagger}$ is negative definite but $f$ is also
negative (because $T<0$ in an amplifying system), so $P/P_{\rm Poisson}$ is
still $>1$.

We will review the derivation of the optical shot-noise formula
(\ref{Pexcess}), and the application to absorbing and amplifying disordered
waveguides. The amplifying case is of particular interest in view of the recent
experiments on random lasers,\cite{Wie97,Cao99} which are amplifying media in
which the feedback required for a laser threshold is provided by scattering
from disorder rather than by mirrors.

\section{Optical Shot-Noise Formula}
\begin{figure}[tb]
\hspace*{\fill}
\epsfxsize=10cm
\epsffile{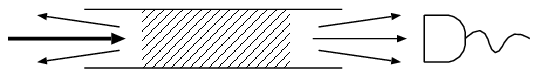}
\hspace*{\fill}
\fcaption{
Coherent light (thick arrow) is incident on an absorbing or amplifying medium
(shaded),
embedded in a waveguide. The transmitted radiation is measured by a
photodetector.
}
\label{aufbau}
\end{figure}

\noindent
In this section we summarize the scattering formulation of the photodection
problem,\cite{Bee98} and derive the formula (\ref{Pexcess}) for the excess
noise.\cite{Pat99} We consider an absorbing or amplifying disordered medium
embedded in a waveguide that supports $N(\omega)$ propagating modes at
frequency $\omega$ (see Fig.\ \ref{aufbau}). The absorbing medium is in thermal
equilibrium at temperature $T>0$. In the amplifying medium, the amplification
could be due to stimulated
emission by an inverted atomic population or to stimulated Raman
scattering.\cite{Hen96} A negative temperature $T<0$ describes
the degree of population inversion in the first case or the density of
the material excitation in the second case.\cite{Jef93} A
complete population inversion or vanishing density corresponds to the
limit $T\rightarrow 0$ from below. The Bose-Einstein function $f(\omega,T)$ is
$\mbox{}>0$ for $T>0$ and $\mbox{}<-1$ for $T<0$.\footnote{
The quantity $f(\omega,T)$ is called the ``population inversion factor'' in the
laser literature, because if $\omega$ is close to the laser frequency $\Omega$
one can express $f=(N_{\rm lower}/N_{\rm upper}-1)^{-1}$ in terms of the ratio
$N_{\rm lower}/N_{\rm upper}=\exp(\hbar\Omega/kT)$ of the population of the
lower and upper atomic levels, with $f=-1$ corresponding to a complete
population inversion.}
\ The absorption or amplification rate $1/\tau_{\rm a}=\omega |\varepsilon''|$
is obtained from the imaginary part $\varepsilon''$
of the (relative) dielectric constant ($\varepsilon''>0$ for absorption,
$\varepsilon''<0$ for amplification).
Disorder causes multiple scattering with rate $1/\tau_{\rm s}$ and
(transport) mean free path $l=c\tau_{\rm s}$ (with $c$ the velocity of light in
the medium). The diffusion constant is $D=\frac{1}{3}cl$. The absorption or
amplification length is defined by $\xi_{\rm a}=\sqrt{D\tau_{\rm a}}$.

The waveguide is illuminated from one end by monochromatic radiation
(frequency $\omega_{0}$, mean photocurrent $I_{0}$) in a coherent state. For
simplicity, we assume that the illumination is in a single
propagating mode (labelled $m_{0}$). At the other end of the waveguide,
a photodetector detects the outcoming radiation. We assume, again for
simplicity, that all $N$ outgoing modes are detected with unit quantum
efficiency. We denote by $p(n)$ the probability to count $n$ photons within a
time
$t$. Its first two moments determine the mean photocurrent $\bar{I}$ and
the noise power $P$, according to\footnote{This definition of $P$ is
equi\-val\-ent to
Eq.\ (\protect\ref{Pdef}); In some papers the noise power is defined with an
extra factor of 2.}
\begin{equation}
\bar{I} = \frac{1}{t} \bar{n}, \qquad
P = \lim_{t\to\infty} \frac{1}{t}\left(
\overline{n^2}-\bar{n}^2\right).
	\label{groessen}
\end{equation}

The outgoing radiation in mode $n$ is described by an annihilation
operator $a_n^{\rm out}(\omega)$, using the convention that modes
$1,2,\ldots,N$ are on the left-hand-side of the medium and modes
$N+1,\ldots,2 N$ are on the right-hand-side. The vector $a^{\rm out}$
consists of the operators $a_1^{\rm out},a_2^{\rm out},\ldots,a_{2
N}^{\rm out}$. Similarly, we define a vector $a^{\rm in}$ for incoming
radiation. These two sets of operators each satisfy the bosonic
commutation relations
\begin{equation}
[ a_n(\omega),a_m^\dagger(\omega')]=\delta_{nm}
	\delta(\omega-\omega'),\;\;\
{[} a_n(\omega),a_m(\omega')] =0, \label{commrel}
\end{equation}
and are related by the
input-output relations\cite{Jef93,Mat95,Grun96}
\begin{equation}
        a^{\rm out}= Sa^{\rm in}+ Ub+
        	Vc^\dagger.
        \label{basiceq}
\end{equation}
We have introduced
the $2N\times2N$ scattering matrix $S$, the $2N\times2N$ matrices
$U,V$, and the vectors $b,c$ of $2 N$ bosonic operators. The reflection and
transmission matrices are $N\times N$ submatrices of $S$,
\begin{equation}
S=\left(\begin{array}{cc}r'&t'\\t&r\end{array}\right).\label{Sdef}
\end{equation}

The operators $b,c$ account for vacuum fluctuations. In order for these
operators to satisfy the bosonic commutation relations (\ref{commrel}), it is
necessary that
\begin{equation}
	U U^\dagger-VV^{\dagger} = \openone - S S^\dagger. 	\label{vvss}
\end{equation}
In an absorbing medium $c\equiv 0$ and $b$ has the expectation value
\begin{equation}
\langle b_{n}^{\dagger}(\omega) b_m(\omega')\rangle
	=\delta_{nm} \delta(\omega-\omega') f(\omega,T),\;\;T>0.
	\label{bexpval}
\end{equation}
Conversely, in an amplifying medium $b\equiv 0$ and $c$ has the expectation
value
\begin{equation}
\langle c_{n}(\omega) c^{\dagger}_m(\omega')\rangle
	=-\delta_{nm} \delta(\omega-\omega') f(\omega,T),\;\;T<0.
	\label{cexpval}
\end{equation}

The probability $p(n)$ that $n$ photons are counted in a time $t$ is given
by\cite{Man95}
\begin{equation}
p(n) = \frac{1}{n!} \langle : W^n e^{-W} : \rangle,
\end{equation}
where the
colons denote normal ordering with respect to $a^{\rm out}$, and
\begin{eqnarray}
&&  W = \int_0^t
	dt' \sum_{n=N+1}^{2N} a_n^{{\rm out}\dagger}(t')
        	a_n^{\rm out}(t'),
        \label{Wanfang} \\
&&  a^{\rm out}_n(t)=(2\pi)^{-1/2} \int_0^\infty d\omega\,
        		e^{-i\omega t} a^{\rm out}_n(\omega). \label{avont}
\end{eqnarray}
Expectation values of a normally ordered
expression are readily computed using the optical equivalence
theorem.\cite{Man95} Application of this theorem to our problem
consists in discretising the frequency in infinitesimally small steps
of $\Delta$ (so that $\omega_p = p \Delta$) and then replacing the
annihilation operators $a^{\rm in}_n(\omega_p)$, $b_n(\omega_p)$,
$c_n(\omega_p)$ by complex
numbers $a^{\rm in}_{np}$, $b_{np}$, $c_{np}$. The coherent state of the
incident
radiation corresponds to a non-fluctuating value of $a_{np}^{\rm in}$,
such that $|a_{np}^{\rm in}|^2=\delta_{nm_{0}}\delta_{p p_0} 2 \pi I_{0} /
\Delta$ (with $\omega_0=p_0\Delta$).
The thermal
state of the vacuum fluctuations corresponds to uncorrelated Gaussian
distributions of the real and imaginary parts of the numbers $b_{np}$ and
$c_{np}$,
with zero mean and variance $\langle |b_{np}|^2
\rangle =-\langle |c_{np}|^2 \rangle=f(\omega_p,T)$, in accordance with Eqs.\
(\ref{bexpval}) and (\ref{cexpval}).

To evaluate the moments of the photocount distribution we need to perform
Gaussian averages. The first two moments determine $\bar{I}$ and $P$. The
results are\cite{Bee98,Pat99}
\begin{eqnarray}
\bar{I} &=& I_{0}
        [t^\dagger t]_{m_{0}m_{0}}
+\int_{0}^{\infty}\frac{d\omega}{2\pi}\,f(\omega,T){\rm Tr}\,(\openone-r
r^\dagger - t t^\dagger ),\label{Ibarresult}\\
P&=&\bar{I} +  2
        I_{0}f(\omega_{0},T)
        [t^\dagger
        (\openone-r r^\dagger - t t^\dagger )
	t]_{m_{0}m_{0}}\nonumber\\
&&\mbox{}+\int_{0}^{\infty}\frac{d\omega}{2\pi}\,f(\omega,T)^{2}\,{\rm
Tr}\,(\openone-r r^\dagger - t t^\dagger )^{2}.\label{Presult}
\end{eqnarray}
The mean photocurrent is the sum of two terms, a term $\propto I_{0}$ equal to
the transmitted part of the incident current and a term $\propto f$ that
represents the thermal emission of radiation. The noise power is the sum of
three terms, the Poisson noise $\bar{I}$ plus two sources of excess noise. The
term $\propto f^{2}$ is due to thermal emission while the term $\propto I_{0}f$
is the excess noise due to beating of vacuum fluctuations with the incident
radiation. For a unitary scattering matrix both terms vanish and $P=\bar{I}$
equals the Poisson value.

The contributions from thermal emission to $\bar{I}$ and $P$ can be eliminated
by filtering the output through a narrow frequency window around $\omega_{0}$.
Only the terms proportional to the incident current $I_{0}$ remain,
\begin{eqnarray}
\bar{I} &=&I_{0}
        [t^\dagger t]_{m_{0}m_{0}},\label{Ibarresult2}\\
P&=&\bar{I} +  2
        I_{0}f(\omega_{0},T) [t^\dagger
        (\openone-r r^\dagger - t t^\dagger )t]_{m_{0}m_{0}}.\label{Presult2}
\end{eqnarray}
This yields the optical shot noise formula (\ref{Pexcess}) discussed in the
introduction.

\section{Absorbing Random Medium}
\noindent
We consider an ensemble of absorbing disordered waveguides, with different
realizations of the disorder, and evaluate the ensemble averages of Eqs.\
(\ref{Ibarresult2}) and (\ref{Presult2}). For a random medium the dependence on
the index $m_{0}$ of the incident radiation is insignificant on average, so we
may replace the average of a matrix element $[\cdots]_{m_{0}m_{0}}$ by the
average of the normalized trace $N^{-1}{\rm Tr}$. Moments of $rr^\dagger$ and
$tt^\dagger$ in the presence of absorption have been computed by
Brouwer\cite{Bro98} using the methods of random-matrix theory, in the regime
that both the length $L$ of the waveguide and the absorption length $\xi_{\rm
a}$ are much greater than the mean free path $l$ but much less than the
localization length $Nl$. This is the large-$N$ regime $N\gg L/l,\xi_{\rm
a}/l\gg 1$. The ratio $L/\xi_{\rm a}\equiv s$ is arbitrary.

\begin{figure}[tb]
\hspace*{\fill}
\epsfxsize=10cm
\epsffile{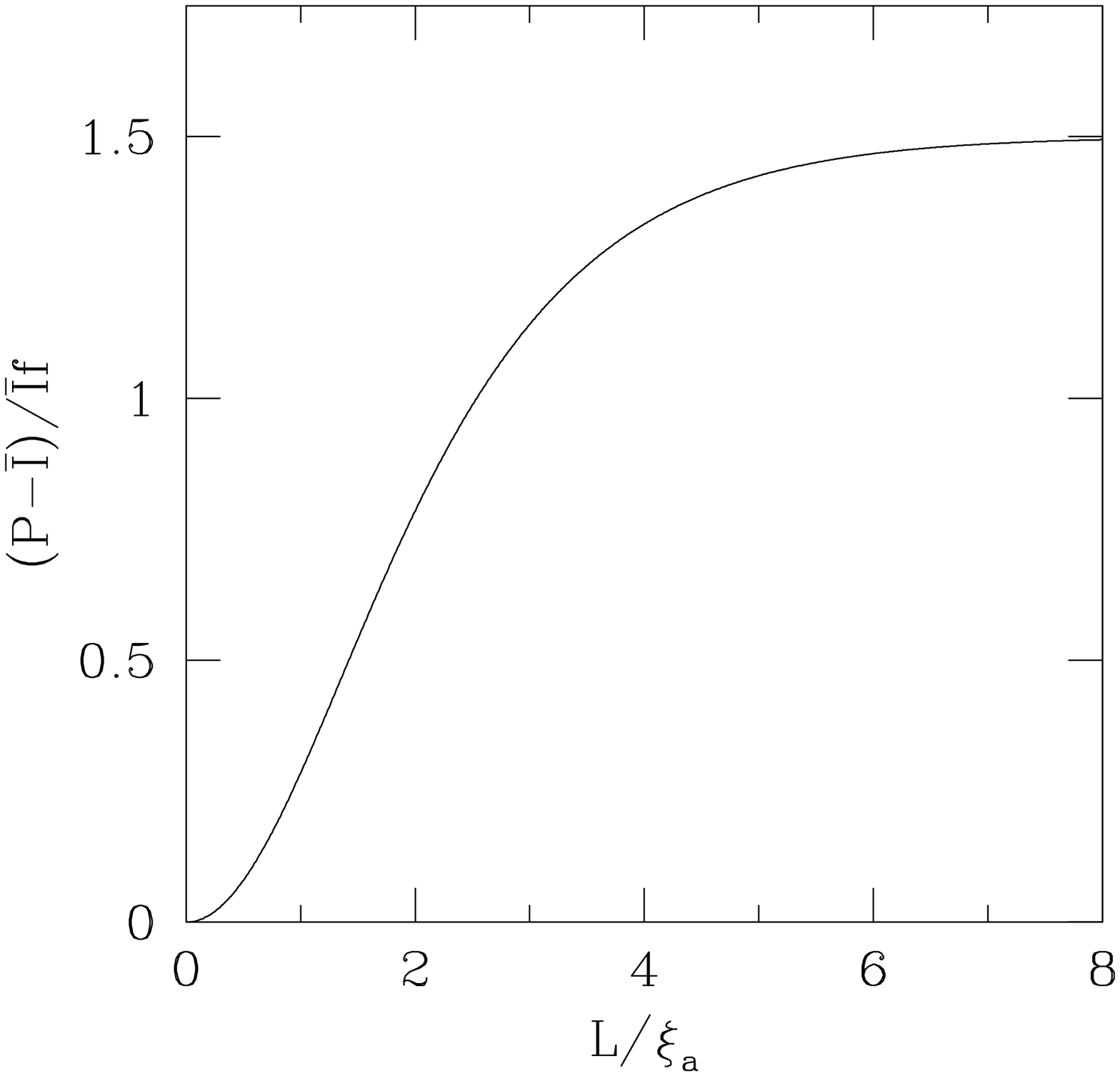}
\hspace*{\fill}
\fcaption{
Excess noise power for an absorbing disordered waveguide, computed from Eqs.\
(\protect\ref{barIslabab}) and (\protect\ref{Pslabab}). The ratio $P/\bar{I}$
tends to $1+\frac{3}{2}f$ for $L\gg\xi_{\rm a}$.
}
\label{slababs}
\end{figure}

The result is\cite{Pat99}
\begin{eqnarray}
\bar{I} & = &
        \frac{4l}{3 L}I_{0}
        \frac{s}{\sinh s},\label{barIslabab} \\
P & = &\bar{I}+
        \frac{2l}{3 L} I_{0}f s\left[
        \frac{3}{\sinh s}-\frac{2 s + {\rm cotanh}\, s}{\sinh^2 s}
     - \frac{s\,{\rm cotanh}\, - 1}{\sinh^3 s}
	+ \frac{s}{\sinh^4 s}
        \right].\label{Pslabab}
\end{eqnarray}
The ratio $P/P_{\rm Poisson}\equiv P/\bar{I}$ increases from $1$ to
$1+\frac{3}{2}f$ with increasing $s$, see Fig.\ \ref{slababs}. The limiting
value $P/P_{\rm Poisson}\rightarrow 1+\frac{3}{2}f(\omega_{0},T)$ for
$L\gg\xi_{\rm a}$ depends on temperature and frequency through the
Bose-Einstein function, but is independent of the scattering or absorption
rates. This might be seen as the optical analogue of the universal limiting
value $P/P_{\rm Poisson}\rightarrow\frac{1}{3}$ for $L\gg l$ of the electronic
shot noise.\cite{Bee92,Nag92}

\begin{figure}[tb]
\hspace*{\fill}
\epsfxsize=10cm
\epsffile{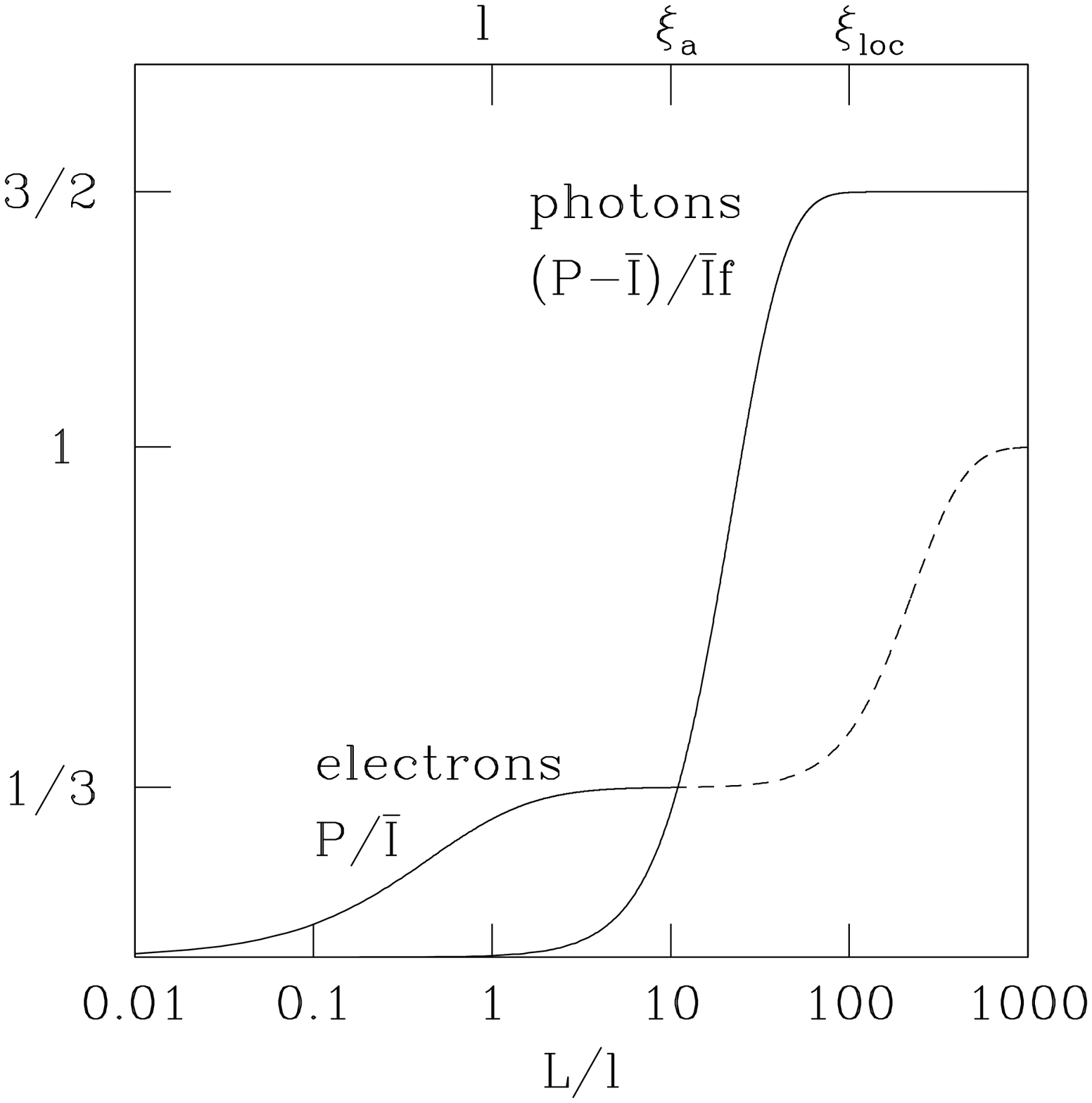}
\hspace*{\fill}
\fcaption{
Different dependence for electrons and photons of the shot-noise power $P$ on
the length $L$ of the waveguide. The curve for photons is the same as in Fig.\
\protect\ref{slababs}, the solid curve for electrons has been calculated in
Ref.\ 25, the dashed curve is a qualitative interpolation. We
have assumed a factor of 10 between the mean free path $l$, the absorption
length $\xi_{\rm a}$, and the localization length $\xi_{\rm loc}=Nl$. (For
electrons, the absorption length should be ignored.)
The electronic $P$ increases from $0$ to $\frac{1}{3}$ of the Poisson value
$\bar{I}$ when $L$ becomes larger than $l$, and then increases further to full
Poisson noise at $\xi_{\rm loc}=Nl$. The photonic
$P$ has only a single transition, at $\xi_{\rm a}$, from $\bar{I}$ to
$(1+\frac{3}{2})\bar{I}$. Nothing happens at $L=l$ or $L=\xi_{\rm loc}$ to the
shot noise of coherent radiation.
}
\label{ep}
\end{figure}

In the derivation of the limiting value of $P/\bar{I}$ we have assumed that the
length $L$ of the waveguide remains small compared to the localization length
$\xi_{\rm loc}=Nl$. What happens in the localized regime $L\gtequiv \xi_{\rm
loc}\,$? Using the results from Ref.\ 24 we find that the ensemble
averages $\langle\bar{I}\rangle$ and $\langle P\rangle$ of current and noise in
the localized regime are suppressed below the results Eqs.\ (\ref{barIslabab})
and (\ref{Pslabab}) in the diffusive regime,\footnote{
The precise result is $\langle\bar{I}\rangle=(1+\frac{3}{2}f)^{-1}\langle
P\rangle=\frac{8}{3}(l/\xi_{\rm a})I_{0}\exp(-L/\xi_{\rm a}-\frac{3}{8}L/Nl)$
in the regime $N,L/l\gg\xi_{\rm a}/l\gg 1$, for any value of $L/Nl\ltequiv
Nl/\xi_{\rm a}$.}
\ but the ratio $\langle P\rangle/\langle\bar{I}\rangle$ remains equal to
$1+\frac{3}{2}f$. This is a remarkable difference with the electronic analogue,
where $P$ becomes equal to the Poisson noise $\bar{I}$ in the localized regime.
The difference between shot noise for electrons and photons is summarized in
Fig.\ \ref{ep}.

\section{Amplifying Random Medium}
\begin{figure}[tb]
\hspace*{\fill}
\epsfxsize=10cm
\epsffile{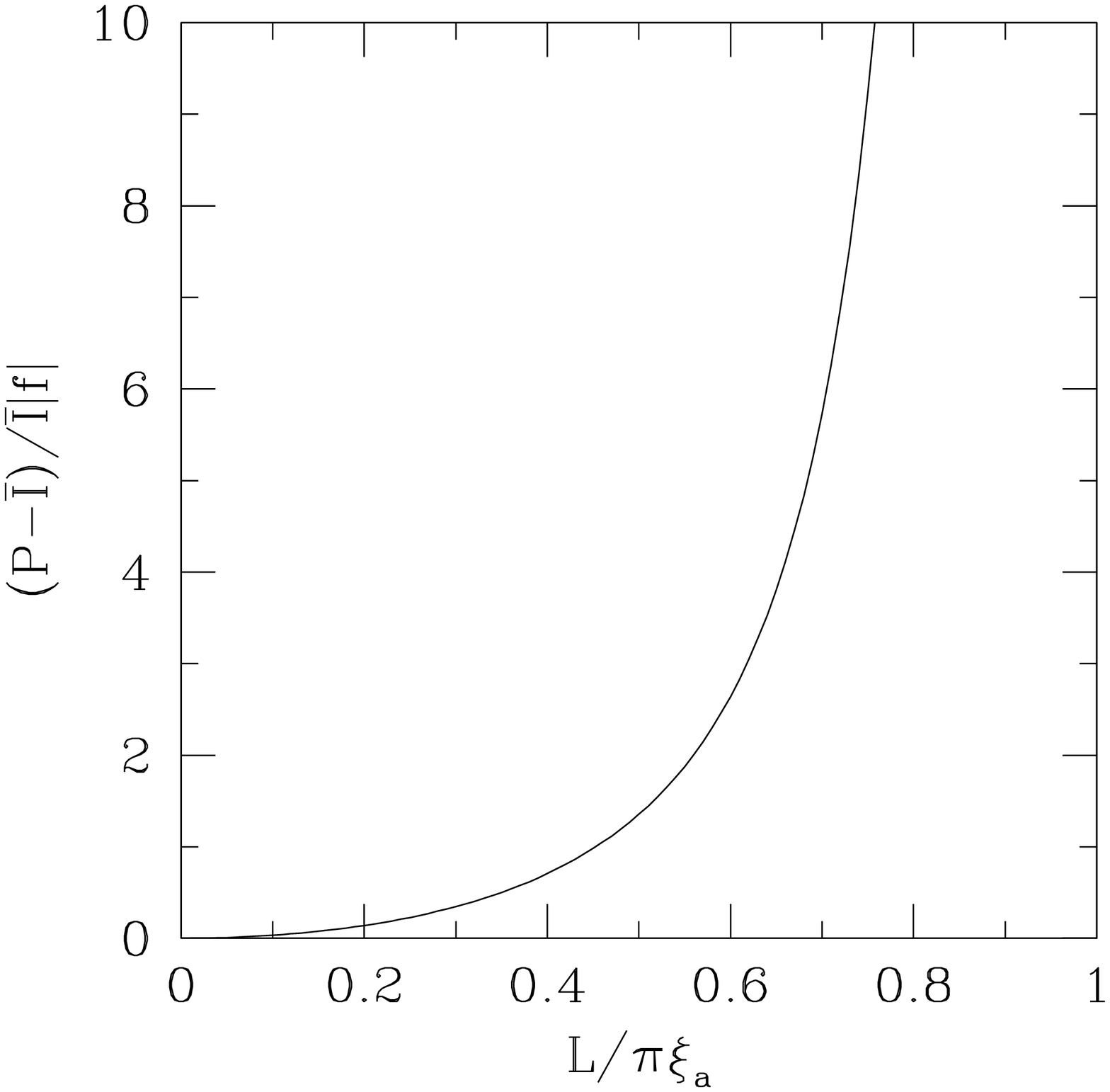}
\hspace*{\fill}
\fcaption{
Excess noise power for an amplifying disordered waveguide, computed from Eqs.\
(\protect\ref{barIslabam}) and (\protect\ref{Pslabam}). The ratio $P/\bar{I}$
diverges at the laser threshold $L=\pi\xi_{\rm a}$.
}
\label{slabamp}
\end{figure}

\noindent
The results for an amplifying disordered waveguide in the large-$N$ regime
follow from Eqs.\ (\ref{barIslabab}) and (\ref{Pslabab}) for the absorbing case
by the substitution $\tau_{\rm a}\rightarrow -\tau_{\rm a}$, or equivalenty
$s\rightarrow is$. One finds
\begin{eqnarray}
\bar{I}& = & \frac{4l}{3 L} I_{0}
        \frac{s}{\sin s},\label{barIslabam} \\
P & = &\bar{I}+
        \frac{2 l}{3 L} I_{0}f s \left[
        \frac{3}{\sin s}-\frac{2 s - {\rm cotan}\, s}{\sin^2 s}
      + \frac{s\,{\rm cotan}\, s - 1}{\sin^3 s}
        - \frac{s}{\sin^4 s}
        \right].\label{Pslabam}
\end{eqnarray}
Recall that the Bose-Einstein function $f<-1$ in an amplifying medium. As shown
in Fig.\ \ref{slabamp}, the ratio $P/\bar{I}$ increases without bound as the
length $L\rightarrow\pi\xi_{\rm a}$ or, equivalently, the amplification rate
$1/\tau_{\rm a}\rightarrow\pi^{2}D/L^{2}$. This is the laser threshold.

To understand better the behavior close to the laser threshold, we consider the
scattering matrix $S(\omega)$ as a function of complex frequency $\omega$. In
the absence of amplification all poles (resonances) of $S$ are in the lower
half of the complex plane, as required by causality. Amplification shifts the
poles upwards by an amount $1/2\tau_{\rm a}$. The laser threshold is reached
when the first pole hits the real axis, say at resonance frequency $\Omega$.
For $\omega$ near $\Omega$ the scattering matrix has the generic form
\begin{equation}
	S_{nm} = \frac{\sigma_n
	\sigma_m}{\omega-\Omega+\case{1}{2} i \Gamma -
	i/2\tau_{\rm a}},
	\label{tempeq11}
\end{equation}
where $\sigma_n$
is the complex coupling constant of the resonance to
the $n$-th mode in the waveguide and $\Gamma$ is the decay rate. The laser
threshold is at $\Gamma\tau_{\rm a}=1$. We will now show that, while $P$ and
$\bar{I}$ diverge at the laser threshold, the signal-to-noise ratio ${\cal
S}=\bar{I}^{2}/P$ has a finite limit --- independent of $\sigma_{n},\Gamma$, or
$\tau_{\rm a}$.\cite{Pat99}

We assume that the incident radiation has frequency $\omega_0=\Omega$.
Substitution of Eq.\ (\ref{tempeq11}) into Eqs.\ (\ref{barIslabam}) and
(\ref{Pslabam}) gives the simple result
\begin{equation}
{\cal S}=\frac{I_{0}|\sigma_{m_{0}}|^2}{2|f|\Sigma}, \;\;
	\Sigma = \sum_{n=1}^{2 N} |\sigma_n|^2.
	\label{noiseallgemein}
\end{equation}
The total coupling constant $\Sigma=\Sigma_{\rm l}+\Sigma_{\rm r}$ is the sum
of the coupling constant $\Sigma_{\rm l}=\sum_{n=1}^{N} |\sigma_n|^2$ to the
left end of the
waveguide and the coupling constant $\Sigma_{\rm r}=\sum_{n=N+1}^{2 N}
|\sigma_n|^2$
to the right. The ensemble average $\langle|\sigma_{m_{0}}|^{2}/\Sigma\rangle$
is independent of $m_{0}\in[1,N]$, hence
\begin{equation}
	\langle{\cal S}\rangle = \frac{I_{0}}{2|f|N}
	\langle \Sigma_{\rm l} / \Sigma \rangle= \frac{I_{0}}{4|f| N}
	,\label{Sresult}
\end{equation}
since $\langle\Sigma_{\rm l}/\Sigma\rangle=\langle\Sigma_{\rm
r}/\Sigma\rangle\Rightarrow\langle\Sigma_{\rm l}/\Sigma\rangle=1/2$. The
signal-to-noise ratio of the incident coherent radiation (with noise power
$P_{0}=I_{0}$) is given by ${\cal S}_{0}=I_{0}^{2}/P_{0}=I_{0}$. The ratio
${\cal S}/{\cal S}_{0}$ is the reciprocal of the noise figure of the amplifier.
The signal-to-noise ratio of the transmitted radiation is maximal for complete
population inversion, when $|f|=1$ and $\langle{\cal S}\rangle$ is smaller than
${\cal S}_{0}$ by a factor $4N$. This universal limit $\langle{\cal S}/{\cal
S}_{0}\rangle\rightarrow 1/4N$ does not require large $N$, but holds for any
$N=1,2,\ldots$. It is the multi-mode generalization of a theorem for the
minimal noise figure of a single-mode linear amplifier.\cite{Hen96,Cav82}

\section{Outlook}
\noindent
We conclude by mentioning some directions for future research. In the
electronic case it is known that the result $P/\bar{I}=1/3$ for the Fano factor
of a diffusive conductor can be either computed from the scattering
matrix\cite{Bee92} (using random-matrix theory) or from a kinetic equation
known as the Boltzmann-Langevin equation.\cite{Nag92} Here we have shown using
the former approach that the optical analogue is a Fano factor of
$1+\frac{3}{2}f$ for a disordered waveguide longer than the absorption length.
To obtain this result from a kinetic equation one needs a Boltzmann-Langevin
equation for bosons. Work in this direction is in progress.\cite{Mis99}

The effect of localization on the Fano factor is strikingly different for
electrons and photons. In the electronic case the average $\langle
P/\bar{I}\rangle$ goes to $1$ in the localized regime, but we have found for
the optical case that the ratio $\langle P\rangle/\langle\bar{I}\rangle$ of
average noise and average current is unchanged as the length of the waveguide
becomes longer than the localization length. We surmise that the same applies
to the average $\langle P/\bar{I}\rangle$ of the ratio, but this remains to be
verified. (In the diffusive regime the difference between the two averages can
be neglected.)

In the case of an amplifying disordered waveguide we have restricted ourselves
to the linear regime below the laser threshold. Above threshold the
fluctuations in the amplitude of the electromagnetic field are strongly
suppressed and only phase fluctuations remain.\cite{Man95} These determine the
quantum-limited linewidth of the radiation. A theory for this linewidth in a
random medium is under development.\cite{Pat00} The application to a disordered
waveguide would require a knowledge of the statistics of the poles of the
scattering matrix in such a system, which is currently lacking.\footnote{
Since in Sec.\ 4 the laser threshold was found to be at $1/\tau_{\rm
a}=\pi^{2}D/L^{2}$ in the large-$N$ limit, we conclude that
$\Gamma=\pi^{2}D/L^{2}$ is the minimal decay rate in that limit. In other
words, the density of $S$-matrix poles for a disordered waveguide without
amplification should vanish for ${\rm Im}\,\omega\,>-\pi^{2}D/2L^{2}$ if
$N\rightarrow\infty$. This density is unknown, but a similar gap in the density
of poles has been found for the scattering matrix of a chaotic
cavity.\protect\cite{Fyo97}}

The recent interest in the Hanbury-Brown and Twiss experiment for electrons in
a disordered metal\cite{But99} suggests a study of the optical case. The
formalism presented here for auto-correlations of the photocurrent can be
readily extended to cross-correlations,\cite{Pat98} but it has not yet been
applied to a random medium.

We do not know of any experiments on photon shot noise in a random medium, and
hope that the theoretical predictions reviewed here will stimulate work in this
direction. The universal limits of the Fano factor in the absorbing case and
the signal-to-noise ratio in the amplifying case seem particularly promising
for an experimental study.

\nonumsection{Acknowledgements}
\noindent
We have benefitted from discussions with P. W. Brouwer, E. G. Mishchenko, and
H. Schomerus.
This work was supported by the ``Ne\-der\-land\-se
or\-ga\-ni\-sa\-tie voor We\-ten\-schap\-pe\-lijk On\-der\-zoek'' (NWO)
and by the ``Stich\-ting voor Fun\-da\-men\-teel On\-der\-zoek der
Ma\-te\-rie'' (FOM).

\nonumsection{References}

\end{document}